\documentclass[11pt]{article}

\usepackage{style_file}
\usepackage{authblk}
\usepackage{xcolor}
\newcommand{\titletext}{Bond strength uncertainty quantification via confidence intervals for nondestructive evaluation of bonded composites}

\title{\titletext}

\author[1]{Michael C. Stanley}
\author[2]{Peter W. Spaeth}
\author[2]{James E. Warner}
\author[2]{Matthew R. Webster}
\affil[1]{Analytical Mechanics Associates, Hampton, VA 23666, USA.}
\affil[2]{NASA Langley Research Center, Hampton, VA 23681, USA.}

\date{\today}

\begin{document}

\maketitle

\begin{abstract}
As bonded composite materials are used more frequently for aerospace applications, it is necessary to certify that parts achieve desired levels of certain physical characteristics (e.g., strength) for safety and performance.
Nondestructive evaluation (NDE) of adhesively bonded structures enables verification of bond physical characteristics, but uncertainty quantification (UQ) of NDE estimates is crucial for understanding risks, especially for NDE estimates like bond strength.
To address the critical need for NDE UQ for adhesive bond strength estimates, we propose an optimization--based approach to computing finite--sample confidence intervals showing the range of bond strengths that could feasibly be produced by the observed data.
A statistical inverse model approach is used to compute a confidence interval of specimen interfacial stiffness from swept--frequency ultrasonic phase observations and a method for propagating the interval to bond strength via a known interfacial stiffness regression is proposed.
This approach requires innovating the optimization--based confidence interval to handle both a nonlinear forward model and unknown variance and developing a calibration approach to ensure that the final bond strength interval achieves at least the desired coverage level.
Using model assumptions in line with current literature, we demonstrate our approach on simulated measurement data using a variety of low to high noise settings under two prototypical parameter settings.
Relative to a baseline approach, we show that our method achieves better coverage and smaller intervals in high--noise settings and when a nuisance parameter is near the constraint boundary.
% Using model assumptions in line with current literaure, we demonstrate our approach on experimental data collected on glass adherends bonded with UV-cured optical adhesive to show that our approach generates better calibrated and often tighter confidence intervals compared to a baseline.
\end{abstract}

\section{Introduction} \label{sec:intro}
% what is the overall goal and why do we have it?
While next--generation aircraft will increasingly rely upon bonded composite structures, there is no widely--adopted nondestructive evaluation (NDE) method to evaluate adhesive bond strength.
The strength of an adhesive bond between two composites is influenced by many factors.  Among these factors are surface preparation in the adherends, accidental contamination of the bonding surface, and processing parameters used in the adhesive cure.  
The strength of the adhesive joint in service could further be affected by environmental aging and repeated mechanical loading.  
Due to this potential for uncertainty in bond strength, it is necessary to  develop nondestructive techniques to certify bonded structures before and during use.

% Adding some background context for other work that has been done on ultrasonic assessment of bond strength
Some promising approaches for detecting weak adhesive joints have been developed that focus on assessing the mechanical bonding between adhesives and adherents.  
A modeling framework considering the adhesive bond as a network of massless springs was developed by \cite{tattersall1973} and generalized for finite--thickness adhesives by \cite{baikthompson1984}.  
Cantrell analyzed the atomic-scale physics of adhesive bonding (e.g., \citep{cantrell2015hydrogen}) and showed that the interfacial stiffness of the spring network models could be related to the number of bonds per unit area in the adhesive region, which is directly linked to bond strength.

Experimental methods for measuring the interfacial stiffness associated with spring network models include various forms of angle beam ultrasonic spectroscopy (ABUS) \citep{cantrell2004determination,lavrentyev1997determination} and characteristic frequency assessments of compressional and shear wave modes \citep{nagy1991ultrasonic}, among others.  
A phase--based approach by \cite{haldren2019} simplified the experimental setup relative to ABUS and improved the signal-to-noise ratio relative to characteristic frequency assessments.  
Haldren used swept--frequency ultrasonic phase measurements made on ultraviolet-cured adhesives at varying degrees of curing and coupled these measurements with mechanical testing to statistically model a linear relationship between the interfacial stiffness of the bond and its strength.
In doing so, \cite{haldren2019} primarily sought to move beyond the detection of good and bad bonds and achieve a goal of successfully detecting intermediate bond strengths.
We build upon the work of \cite{haldren2019} and add a layer of uncertainty quantification (UQ) to their approach by developing an approach to compute bond strength confidence intevals using the components described above.
Since a point estimate of a physical quantity like bond strength is meaningless without a characterization of its variability, this UQ layer adds a crucial quantification of the information content of the data.

% what is our approach?
% Continuing
We leverage the work of optimization--based confidence intervals \citep{stark1992, stanley2022, patil2022, batlle2023, stanley2025} to return a set interfacial stiffness values that are consistent with the observed ultrasonic phases.
This approach natively incorporates known physical constraints on the forward model parameters and aims to produce intervals achieving finite--sample (non--asymptotic) coverage of the true underlying interfacial stiffness of each specimen.
\cite{haldren2019} further assume the existence of side--data consisting of pairs of observed interfacial stiffness and bond strength values.
Under their linear model assumptions, we propose a method to propagate the optimized interfacial stiffness confidence interval through the bond strength--interfacial stiffness regression to obtain a confidence interval on bond strength.

% why is our approach reasonable statistically?
The Frequentist statistical paradigm that can provide estimates in the form of confidence intervals is sensible to use in this NDE use--case since an operational UQ method would be deployed a large number of times to certify a large number of parts.
Intuitively, one would like to use a procedure with a guarantee that it works as promised some chosen percent of the time.
A non--asymptotic confidence interval approach achieves this criterion since for a chosen confidence level $1 - \alpha$ (e.g., $0.95$), we can be sure that our interval is covering the true physical quantity of interest at least $(1 - \alpha) \times 100$\% of the time.
Although statistical inverse problems are typically handled in a Bayesian paradigm involving prior distributions on model parameters \citep{tarantola2005,kaipio2005}, priors are known to bias frequency properties of credible intevals \citep{patil2022} and thus provide no frequency guarantee on the produced interval.
This biasing effect is often the result of the prior smuggling unintended information into the problem \citep{stark2015} which can make the resulting credible intervals over-- or under--conservative depending on the situation.
By contrast, since optimization--based confidence intervals allow the native incorporation of parameter constraints into the interval estimation they are able to leverage physical information in a way that does not affect interval validity.
We provide a clear demonstration of this effect in \Cref{sec:estimating_cov_exp_len}.

% describe the statistical methods advancements made to address this problem
To enhance the appropriateness of optimization--based intervals to this NDE scenario, we propose three methodological innovations.
The existing literature assumes the forward model is linear and the noise variance is known, two unrealistic assumptions for this application.
We relax these assumptions by (i) providing a way to handle the model nonlinearity and (ii) a classically motivated approach to handle the unknown noise variance.
To propagate the finite--sample confidence interval on interfacial stiffness, we (iii) develop a procedure to obtain a confidence interval on bond strength that achieves the desired level of coverage.

% outline of paper
The rest of the paper is structured as follows.
In \Cref{sec:approach} we describe the background modeling assumptions and fundamental ideas required for our approach.
\Cref{sec:ultrasonic_model} describes the data--generating processes for both the ultrasonic phase observations and the observed pairs of interfacial stiffness and bond strength.
\Cref{sec:interval_defs} describes the necessary background on optimization--based confidence intervals and the settings under which they are known to achieve coverage.
\Cref{sec:linearization} provides a quick overview of the linearization we use as part of our approach.
\Cref{sec:method} provides the fine details of our approach, with \Cref{sec:two_step_proc} detailing our nonlinear and unknown variance innovations of the optimization--based confidence interval method, \Cref{sec:ls_baseline} describing a baseline confidence interval construction against which we compare our approach, and \Cref{sec:prop_interval} detailing the modular interval propagation approach.
\Cref{sec:experiment} presents simulated numerical experiments demonstrating the performance of our method.
\Cref{sec:estimating_cov_exp_len} shows how coverage and expected interval length vary as a function of noise variance for two stragically chosen true parameter settings.
\Cref{sec:bond_strength_model_effects} examines the impact of the bond strength model parameters on the final bond strength interval and \Cref{sec:constraint_effects} examines the effect of contracting the known physical constraints to show how extra information via the constraints affects the results.
Finally, \Cref{sec:disc_conc} provides a discussion of our approach and outlines several future directions.

A nearly complete list of the mathematical notation used throughout the paper is presented in \Cref{app:math_symbols} in \Cref{tab:symbols}.
We isolate the definition and use of the original notation from \cite{haldren2019} to \Cref{sec:ultrasonic_model} to preserve the original notation without conflicting with the standard statistical notation used throughout the rest of the paper.
Generally, we denote real quantities with lowercase Roman characters, vectors with bolded lowercase Roman characters, matrices with bolded uppercase Roman characters, parameter vectors with bolded Greek characters, miscoverage levels with Greek characters, and the true values of unknown values with a superscript $*$ symbol.

\section{Background: Modeling assumptions and key concepts} \label{sec:approach}

\subsection{Ultrasonic and bond strength models} \label{sec:ultrasonic_model}

\begin{figure}[h]
    \centering
    \includegraphics[width=0.9\textwidth]{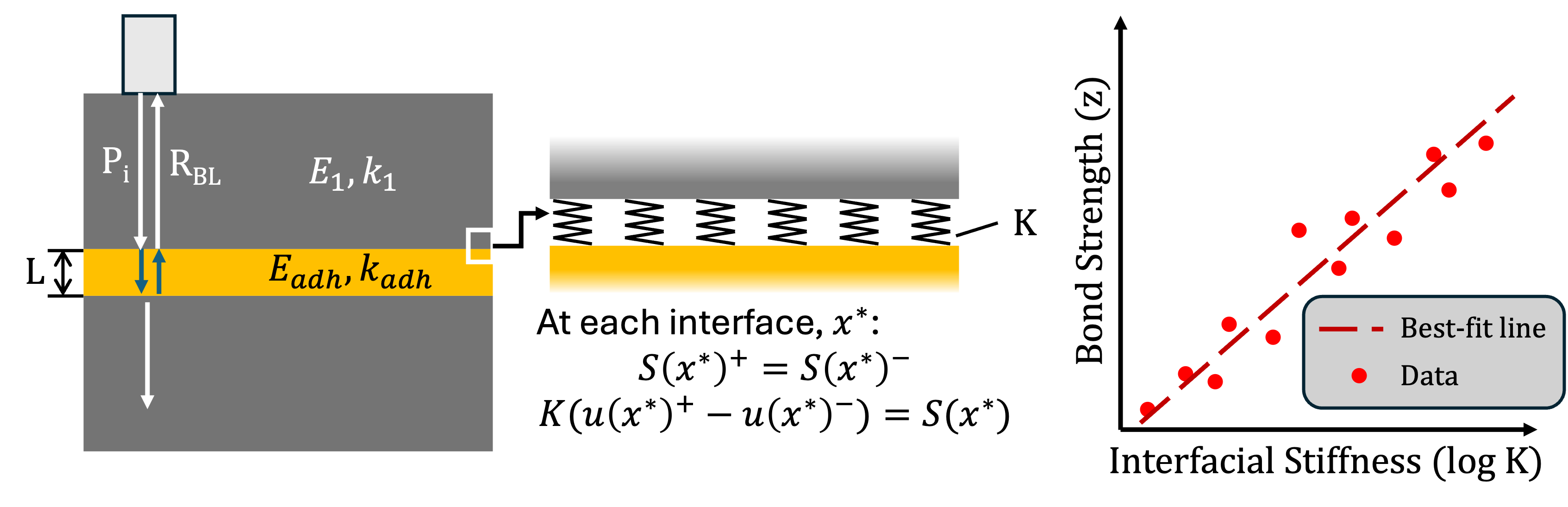}
    \caption{Reproductions of the ultrasonic forward model (\textbf{left}) and bond strength regressed on interfacial stiffness (\textbf{right}) as described in \cite{haldren2019}. Once a parameter vector has been set, the arrows on the left illustrate the direction and magnitude of the ultrasonic waves within the bonded composite. The waves start at the top from the ultrasonic transducer and proceed through the first adherent layer. Some are reflected on the boundary and some pass into the adhesive layer. A similar process occurs going from the adhesive to the second adherent. The red dots on the right illustrate observed interfacial stiffness and bond strength pairs from which a best--fit regression line can be constructed.}
    \label{fig:ultrasonic_diagram}
\end{figure}

% Inserting details of the forward model
The forward model used in \cite{haldren2019} allows us to first consider the statistical inverse problem of inferring interfacial stiffness from the observed noisy ultrasonic phase measurements.  
The forward model considers the frequency-dependent reflection coefficient resulting from the interaction of a normal incidence pressure wave ($P_i$) with a tri--layer adhesive structure (see \Cref{fig:ultrasonic_diagram}).  
Using the modeling framework of \cite{baikthompson1984} and assuming negligible mass contribution, boundary conditions are applied at each interface ($x=x^{*}$) in terms of the normal stress, $S(x)$, and $x$-displacement, $u(x)$, at either side of the interface.
Specifically, the interfaces maintain continuity of stress such that $S(x^{*})^{+} = S(x^{*})^{-}$, and any discontinuity in displacement between either side of the interface is counteracted by a linear spring network having stiffness, $K$ (i.e., the interfacial stiffness), such that we have $K(u(x^{*})^{+}-u(x^{*})^{-})=S(x^{*})$.
For tri-layer joints like those shown in \Cref{fig:ultrasonic_diagram}, there are two such interfacial conditions at $x^{*} = 0$ and $x^{*} = L$, where $L$ is the bond thickness.  
In the simplified case of identical upper and lower bonding conditions ($K(x=0) = K(x=L)=K$), the reflection coefficient can be written as, 

\begin{equation}
\begin{gathered}
	R_{BL} = \frac{C_N \cos(k_{adh} L_{BL}) + iS_N \sin(k_{adh} L_{BL})}{C_D \cos(k_{adh} L_{BL}) + iS_D \sin(k_{adh} L_{BL})} \in \mathbb{C} \nonumber \\
	C_{N}=\frac{G_{1}^2}{K},\quad		C_{D}=2G_{1}G_{adh} \left(1+\frac{G_{1}}{K} \right) \nonumber \\
	S_{N}=G_{1}^2-G_{adh}^2+ \left(\frac{G_{1}G_{adh}}{K} \right)^{2}, \quad  S_{D}=G_{1}^2+G_{adh}^2+\frac{G_{1}G_{adh}^{2}}{K} \left(2+\frac{G_{1}}{K} \right), \\
\end{gathered}
\end{equation}

where $E_{1}$, $k_{1}$, and $E_{adh}$, $k_{adh}$ are the Young's modulus and wave number of the upper adherent and adhesive, respectively, $G_{1}=E_{1}k_{1}$, $G_{adh}=E_{adh}k_{adh}$, and $i=\sqrt{-1}$.
Note that in this expression the wave number is defined as $k_{j}=\omega/c_{Lj}+\alpha_{j}i$, where $\omega$ is the frequency of the incident pressure wave and $c_{Lj}$ and $\alpha_{j}$ are the longitudinal wave speed and attenuation in layer $j$, respectively.

For a fixed set of frequencies ($\bomega \in \mathbb{R}^n$), we assume the data are generated according to the following model,
\begin{equation} \label{eq:dgm}
    \by = f(\btheta^*; \bomega) + \bepsilon, \quad \btheta^* \in \bTheta, \quad \bepsilon \sim \cN(\bm{0}, \sigma^2 \bI_n),
\end{equation}
where $\bTheta \subset \mathbb{R}^p$ ($p = 5$ in this case) and $\by \in \mathbb{R}^n$ are the phase angles (in degrees) resulting from the input frequencies ($\bomega$).
Translated from the equations above, $f(\btheta^*; \omega)_i = \arg(R_{BL}) + (a \omega_i + b)$, where $\arg(\cdot)$ denotes the phase angle of the complex input.
Consistent with \cite{haldren2019}, we assume $\bTheta$ takes the form of a hyperrectangle (i.e., we have box constraints for $\btheta^*$).

\begin{figure}[h]
    \centering
    \includegraphics[width=\textwidth]{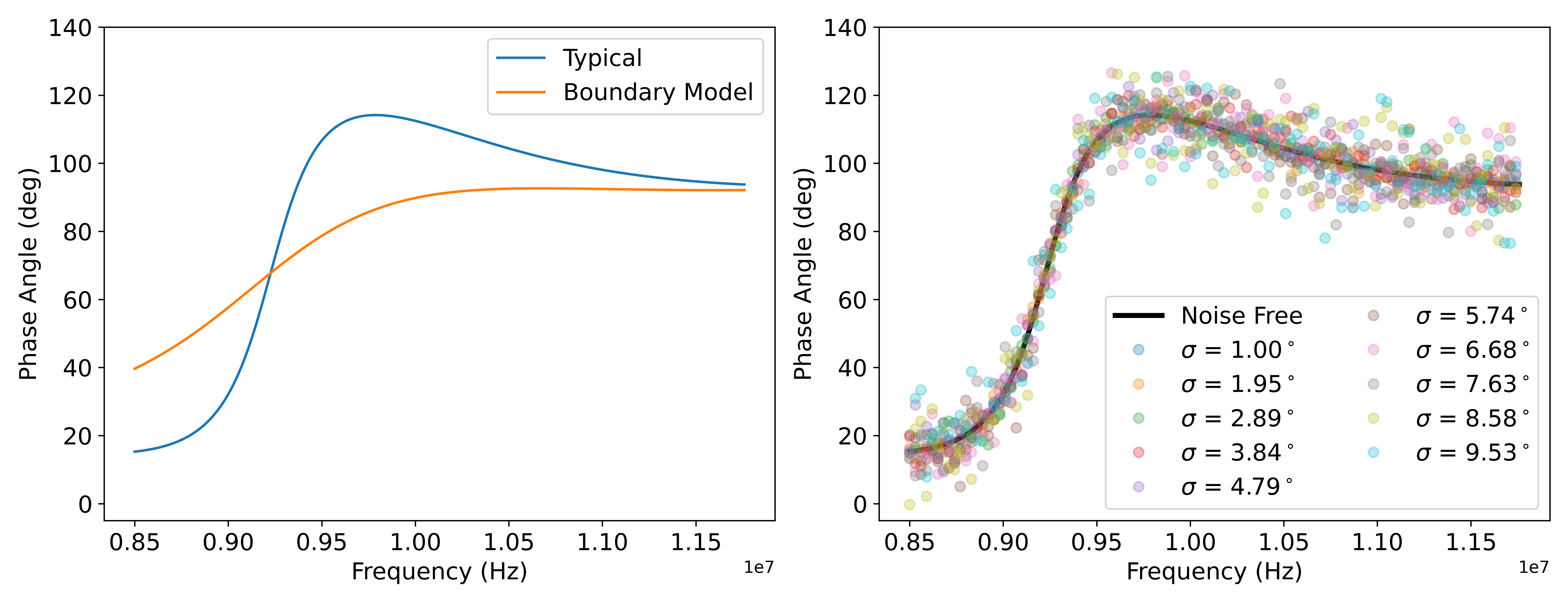}
    \caption{\textbf{(Left)} Noise--free ultrasonic forward model output for a given collection of frequencies, $\bomega$. Two settings of $\btheta^*$ are shown; the ``typical'' setting is far from all parameter constraint boundaries while the ``boundary'' setting sets acoustic attenuation close to its upper bound. The effect of this subtle change is explored in \Cref{sec:estimating_cov_exp_len}. \textbf{(Right)} For the ``typical'' parameter setting, a variety of phase data at different noise levels are generated.}
    \label{fig:data_viz}
\end{figure}

The $\btheta^*$ vector contains five physical components: interfacial stiffness ($\log K_0$), acoustic attenuation ($\alpha_0$), multiplicative correction ($a$), additive correction ($b$), and bond thickness ($L_{BL})$, giving us,
\begin{equation}
    \btheta^* = \begin{pmatrix}
        \log K_0 & \alpha_0 & a & b & L_{BL}
    \end{pmatrix}^T.
\end{equation}
We refer to the $i$th component in this list by $\theta^*_i$.
The $\bomega$ set of frequencies are the ultrasonic frequencies over which the measurements are swept, each of which provides a phase angle output.
\Cref{fig:data_viz} provides a look at some generated data according to \Cref{eq:dgm}, where $\sigma$ is set to the different noise--level standard deviations denoted in the legend.
Under these modeling assumptions, we aim to solve and characterize the uncertainty of an inverse problem of inferring $\log K_0$ from data $\by$.
Inverse problems are often solved in a Bayesian paradigm, i.e., finding a posterior distribution on $\btheta^*$ by setting a prior and likelihood function via the model defined by \Cref{eq:dgm} (e.g., see \cite{kaipio2005, tarantola2005}).
This paper pursues the inverse problem without assuming a prior and thus pursues a non--Bayesian framework where known physical constraints are leveraged to obtain a frequentist confidence interval of a real--valued function of the parameter \citep{stark1992, patil2022, stanley2022}.
We refer to this value as the quantity of interest (QoI).
A frequentist perspective is sensible in the NDE context since an operational method would be used a large number of times thus providing a $1 - \alpha$ confidence interval with a direct physical meaning.

Following Figure 9b in \cite{haldren2019} (and reproduced in the right panel of \Cref{fig:ultrasonic_diagram}), we propose the following statistical data--generating model for bond strength,
\begin{equation} \label{eq:lin_relationship_bs}
    z = \beta_1 \log K_0 + \varepsilon, \quad \varepsilon \sim \mathcal{N}(0, \sigma_B^2).
\end{equation}
We note that \Cref{eq:lin_relationship_bs} does not have an intercept term to maintain consistency with the model in \cite{haldren2019}.
Assuming we have interfacial stiffness and bond strength pairs, $(x_i, z_i)_{i = 1}^N$, we regress bond strength on interfacial stiffness to obtain the maximum likelihood/least--squares estimators, $\hat{\beta}_1$ and $\hat{\beta}_0$ for a fitted model, $z = \hat{\beta}_1 x + \hat{\beta}_0$.

\subsection{Interval definitions} \label{sec:interval_defs}
We wish to compute an interval of interfacial stiffness parameter values consistent with the observed data such that the interval achieves $(1 - \alpha) \times 100$\% coverage. 
Producing an interval with this guarantee says that if the interval construction were used to infer interfacial stiffness, the procedure would contain the true interfacial stiffness $(1 - \alpha) \times 100$\% of the time, regardless of the true unknown parameter vector ($\btheta^*$).
Such a probabilistic guarantee would allow effective NDE to be performed as one could simply check if the interval is on the correct side of a predetermined safety--critical threshold.

For inverse problem UQ with a linear deterministic forward model\footnote{i.e., for a fixed $\bomega \in \mathbb{R}^n$, this would mean there exists a matrix $\bK \in \mathbb{R}^{n \times p}$ such that $f(\btheta; \bomega) = \bK \btheta$} and known parameter constraints, the optimization--based confidence intervals of \cite{batlle2023, patil2022, stanley2022, stanley2025} provide a well--suited interval framework.
For a chosen miscoverage level $\alpha \in (0, 1)$ (e.g., $\alpha = 0.05$), we define an interval by,
\begin{equation} \label{eq:interval_def}
    I_\alpha(\by) := \left[ \min_{\btheta \in D(\by, q)} \bh^T \btheta, \max_{\btheta \in D(\by, q)} \bh^T \btheta \ \right],
\end{equation}
where $D(\by, q) := \left\{\btheta \in \bTheta: \lVert \by - \bK \btheta \rVert_2^2 \leq q \right\}$ and $\bh \in \mathbb{R}^p$ a vector such that $\bh^T \btheta$ is a desired quantity of interest.
In this setting, we can achieve inference on the interfacial stiffness parameter by setting $\bh^T := \begin{pmatrix} 1 & 0 & 0 & 0 & 0 \end{pmatrix}$.
Equivalently, we wish to find a confidence interval for interfacial stiffness while treating the other four inputs as nuisance parameters.
One challenge of using these intervals is selecting $q > 0$ for the chosen $\alpha \in (0, 1)$ such that
\begin{equation} \label{eq:coverage_guarantee}
    \mathbb{P} \left(\bh^T \btheta^* \in I_\alpha(\by) \right) \geq 1 - \alpha, \quad \forall \btheta^* \in \bTheta.
\end{equation}
\Cref{eq:coverage_guarantee} is known as a \textbf{coverage guarantee}.
% Assuming we have a $q$ such that Interval~\eqref{eq:interval_def} achieves $(1 - \alpha) \times 100$\% coverage, the above guarantee says that if the interval construction were used to infer bonded composite interfacial stiffness, the procedure would contain the true interfacial stiffness $(1 - \alpha) \times 100$\% of the time, regardness of the true unknown interfacial stiffness ($\btheta^*$).
% Such a probabilistic guarantee would allow effective NDE to be performed as one could simply check if the interval is on the correct side of a predetermined safety--critical threshold.
We emphasize that Interval~\eqref{eq:interval_def} is a random interval through the randomness of the observation, $\by$, and thus the coverage guarantee makes a statement regarding the frequency with which the random interval covers the fixed unknown true quantity of interest.

\paragraph{Achieving coverage}
Under the Gaussian assumption and known variance $\sigma^2$, there is a simple yet conservative $q$ setting as defined by the ``Simultaneous Strict Bounds'' (SSB) interval in \cite{stanley2022}.
Namely, we know that $\lVert \by - f(\btheta^*; \bomega) \rVert_2^2 \sim \chi^2_{n}$ and thus we can use the quantile $q := \chi^2_{n, \alpha}$ (where $\chi^2_{n, \alpha}$ is the upper--$\alpha$ quantile of a chi--squared distribution with $n$ degrees of freedom) to ensure that $\mathbb{P} \left(\btheta^* \in D(\by, \chi^2_{n, \alpha}) \right) = 1 - \alpha$.
We note that this calibration technique does not require the linear forward model assumption since the calibration is handled in the observation space.
However, the calibration assumes the noise variance is known which we later relax.
% This term ``strict bounds'' originates in \cite{stark1992} in which interval estimators that respected known physical constraints were defined.
% \cite{stanley2022} used this phrase along with the ``simultaneous'' and ``one-at-a-time'' phrasing from \cite{rust_oleary_1994} to denote different coverage objectives to synthesize the terms ``Simultanous Strict Bounds'' and ``One-at-a-time Strict Bounds'' (OSB).
The guarantee in \Cref{eq:coverage_guarantee} follows trivially, albeit possibly leading to over--coverage for the final interval.
The intuitive reason for the SSB conservatism is that the coverage guarantee for the QoI follows from a coverage guarantee on a confidence set for the \emph{entire} parameter vector.
These intervals are called ``simultaneous'' since they achieve the coverage guarantee for \emph{all} possible QoIs.
When the forward model is linear and the forward model matrix has column-rank $p < n$, then setting $q := \min_{\btheta \in \Theta} \lVert \by - \bK \btheta \rVert_2^2 + \chi^2_{p, \alpha}$ defines a valid confidence set in the parameter space.
\footnote{We do not necessarily care about optimizing over a confidence set for the \emph{entire} parameter vector since all but the interfacial stiffness parameter are nuisance parameters.
It is conceivable that one could design a set $D(\by, q)$ such that $\mathbb{P}(\btheta^* \in D(\by, q)) < 1 - \alpha$ but $\mathbb{P} (\bh^T \btheta^* \in I_\alpha(\by)) \geq 1 - \alpha$.
I.e., the set $D(\by, q)$ is not a $1 - \alpha$ confidence set.
This intuition motivates the one--at--a-time strict bounds (OSB) interval in \cite{stanley2022} which endeavors to provide a coverage guarantee for one QoI at a time.
Although we can thematically define this type of interval, it is only known to be valid when the forward model is linear, variance is known, and there are no parameter constraints.
Since this interval is only known to cover in the unconstrained setting, we do not include these intervals in this paper.}

In \Cref{sec:method}, we describe how the simultanous interval construction can be adjusted to accomodate unknown variance along with an approach to accommodate the forward model nonlinearity when solving the endpoint optimizations defining Interval~\eqref{eq:interval_def}.

\subsection{Linearizing the forward model} \label{sec:linearization}
The nonlinear forward model in \Cref{eq:dgm} provides a challenge to both calibrating the coverage and computing Interval~\eqref{eq:interval_def}.
Namely, the endpoint optimizations in Interval~\eqref{eq:interval_def} can become arbitrarily challenging nonlinear programs depending on the forward model.
Part of our proposed strategy to handle the nonlinearity is to linearize the forward model and thus we provide a short note about how that is done here.
Let $\tilde{\btheta} \in \Theta$ be the point around which the forward model is linearized.
In support of this method, we linearize the forward model around the least--squares estimator (see \Cref{sec:two_step_proc} for details) but one can in principle linearize around any point.
Define,
\begin{equation} \label{eq:linearized_model}
    g(\btheta) := \bK(\tilde{\btheta}) \left(\btheta - \tilde{\btheta} \right) + f(\tilde{\btheta}; \bomega),
\end{equation}
where $\bK(\tilde{\btheta}) = \nabla_{\btheta} f(\tilde{\btheta}; \bomega) \in \mathbb{R}^{n \times p}$.
We henceforth equate $\bK \equiv \bK(\tilde{\btheta})$ when the point around which we linearize is clear via context.
Further, define $\bb := - \bK \tilde{\btheta} + f(\tilde{\btheta}, \bomega)$, giving $g(\tilde{\btheta}) = \bK \tilde{\btheta} + \bb$.
Using this linearization, we approximate the data--generating process described by \Cref{eq:dgm} using following Gaussian--linear model,
\begin{equation} \label{eq:linear_model}
    \tilde{\by} = \bK \tilde{\btheta} + \bepsilon,
\end{equation}
where $\tilde{\by} = \by - \bb$ and $\bepsilon \sim N(\bm{0}, \sigma^2 \bI_n)$.

\section{Methodology} \label{sec:method}
In this section we propose a procedure to obtain a non--asymptotic $1 - \alpha$ confidence interval on bond strength by first quantifying the uncertainty of the inverse problem on interfacial stiffness followed by propagating the resulting confidence interval through a confidence band on the relationship between interfacial stiffness and bond strength.
Although Interval~\eqref{eq:interval_def} natively handles known parameter constraints and the variational form of its endponts works well with a computationally defined forward model, its calibration relies upon the linear forward model and known variance assumptions.
We address both of these limitations to apply this method to the inference of interfacial stiffness.
% Additionally, to obtain a bond strength confidence interval we develop an approach to propagate the computed interfacial stiffness interval through a pre--existing regression of bond strength on interfacial stiffness such that the final bond strength interval achieves at least the desired coverage level.
To handle the forward model nonlinearity in the interfacial stiffness inverse problem, in \Cref{sec:two_step_proc} we propose a two--step procedure in which we first linearize the forward model around the constrained nonlinear least--squares estimator followed by computing an optimization--based confidence interval from the linearized inverse problem.
\Cref{sec:two_step_proc} further shows how the unknown variance is estimated via the linearized model residuals and the confidence set critical value is updated from a chi--squared distribution quantile to an $F$--distribution quantile.
\Cref{sec:ls_baseline} then details the baseline against which our method is compared which is a standard statistical approach to constructing a confidence interval that does not natively enforce the known parameter constraints.
Finally, \Cref{sec:prop_interval} provides an approach to propagate the interfacial stiffness interval to obtain a valid bond strength interval.

\subsection{A two--step procedure for interfacial stiffness interval} \label{sec:two_step_proc}
Under the linear forward model and known variance assumptions, there are several known ways to calibrate the optimization--based intervals.
As shown in \Cref{sec:prop_interval}, since we wish to obtain a $1 - \gamma$ confidence interval on interfacial stiffness, we use this confidence level throughout this section.
As mentioned in \Cref{sec:interval_defs}, we can use either $q_1 :=\chi^2_{n, \gamma}$ or $q_2 :=\min_{\btheta \in \Theta} \lVert \by - \bK \btheta \rVert_2^2 + \chi^2_{p, \gamma}$.
If $n \gg p$, it is likely that $q_1 > q_2$.
\footnote{Justification for this intuition can be found in foundational books on the theory of linear regression, e.g., \cite{draper1998, seber2003}.}
Using $q_1$ can be thought of as calibrating in the data space whereas $q_2$ can be thought of as calibrating in the parameter space.
Relaxing the linear assumption, $q_1$ is still valid, since $D(\by, q_1)$ is a valid $1 - \gamma$ confidence set regardless of the structure of the forward model.
However, this confidence set will likely be over--conservative and solving the resulting interval endpoints as defined in Interval~\eqref{eq:interval_def} is numerically challenging.
Thus to reduce conservatism and increase computational tractability, \Cref{alg:two_step_proc} proposes a ``linearize--then--optimize'' process to construct an interfacial stiffness confidence interval with unknown variance.

\begin{algorithm}
\caption{Interfacial Stiffness Interval Estimation Algorithm}
\label{alg:two_step_proc}
\begin{algorithmic}[1]  % The [1] enables line numbering
\Require Miscoverage level $\gamma \in (0, 1)$, vector $\bh \in \mathbb{R}^p$ defining the functional of interest and observed phase data $\by \in \mathbb{R}^n$.
\State Compute the nonlinear least--squares estimator:
\begin{equation}
    \hat{\btheta} := \underset{{\btheta \in \Theta}}{\text{argmin}} \lVert \by - f(\btheta; \bomega) \rVert_2^2
\end{equation}
\State Linearize the nonlinear forward model around $\hat{\btheta}$ by constructing the following objects.
\begin{align}
    \bK &:= \nabla_{\btheta} f\left(\hat{\btheta}; \bomega \right), \\
    \bb &:= -\bK \hat{\btheta} + f\left(\hat{\btheta}; \bomega \right), \\
    \tilde{\by} &:= \by - \bb.
\end{align}
\State Define the feasible set over which the interval endpoints are optimized.
\begin{equation} \label{eq:is_conf_set}
    D(\tilde{\by}, q) := \left\{\btheta \in \Theta: \lVert \tilde{\by} - \bK \btheta \rVert_2^2 \leq q \right\}
\end{equation}
where
\begin{equation}
    q := \min_{\btheta \in \Theta} \lVert \tilde{\by} - \bK \btheta \rVert_2^2 \left(1 + \frac{p}{n - p} F_{\gamma}(p, n - p) \right).
\end{equation}
\State Compute the resulting interval:
\begin{equation}
    I_\gamma(\by) := \left[\min_{\btheta \in D(\tilde{\by}, q)} \bh^T \btheta, \max_{\btheta \in D(\tilde{\by}, q)} \bh^T \btheta \right]
\end{equation}
\Ensure $I_\gamma(\by)$.
\end{algorithmic}
\end{algorithm}

This approach is consistent with the those for nonlinear least--squares as outlined in \cite{bates1988,seber1989, draper1998}, with some results about confidence intervals in nonlinear settings discussed in \cite{donaldson1987} including linearization methods (via different treatments of the covariance matrix), likelihood methods, and lack-of-fit method.
Of these, only the lack-of-fit method produces exact confidence regions at the expense of computational cost with respect to the linearization methods.
% \cite{donaldson1987} define the notion of a solution locus, a subset of the parameter space defined by fixing the forward model at the $n$ observed inputs with the parameters allowed to vary over their bounded space.

If we first solve the nonlinear least--squares problem to find a point around which to linearize the model, we obtain access to a wider range of optimization-based approaches.
The primary mathematical challenge is to show the conditions under which this approach yields a valid interval.
Given a $q$ setting providing coverage for Interval~\eqref{eq:interval_def} when the forward model is linear does not imply coverage for when the nonlinear forward model is linearized around the Maximum Likelihood Estimator (MLE).
Note, under the model assumptions the Maximum Likelihood and least--squares estimators are equivalent.
Intuitively, if the MLE is ``close'' to the true parameter on average and the linearized model around the MLE is not too different from the linearized model around the true parameter, it is reasonable that the coverage guarantee should persist.
\cite{seber1989} provides reasoning that the set defined by \Cref{eq:is_conf_set} is asymptotically the same as the following set,
\begin{equation} \label{eq:non_linearized_set}
    \left\{\btheta \in \Theta: \frac{S(\btheta) - S(\hat{\btheta})}{S(\hat{\btheta})} \leq \frac{p}{n - p} F_{\gamma}(p, n - p) \right\},
\end{equation}
where $S(\btheta) := \lVert \by - f(\btheta; \bomega) \rVert_2^2$.
I.e., \Cref{eq:non_linearized_set} shows the set resulting from \emph{not} linearizing the forward model.
Although \Cref{eq:non_linearized_set} requires less approximation, it is still not clear if this set is valid for all possible nonlinear forward models $f(\cdot)$ in the finite--sample regime.
Although we do not have a mathematical guarantee of the finite--sample validity of \Cref{alg:two_step_proc}, its close relationship to \Cref{eq:non_linearized_set} and the empirical studies in \Cref{sec:estimating_cov_exp_len} provide evidence that it is reasonable for this application where $n$ is typically large.

\paragraph{Handling unknown variance}
The presentation of \Cref{alg:two_step_proc} and its proposed $q$ value differs from the valid $q$ values mentioned in \Cref{sec:interval_defs}.
The current theory and implementation of the optimization-based intervals assumes $\sigma^2$ is known.
Since this assumption is broken in this application, it is necessary to find an alternative way to make $D(\by, q)$ valid.
In this section, we assume the forward model is linear and thus justify the use of the $F$--distribution in \Cref{alg:two_step_proc}.

The strategy of the simultaneous intervals is to first deliver a $1 - \gamma$ confidence set in the parameter space to define the optimization feasible region and then to optimize the QoI over that space including any physical constraints.
Under known variance $\sigma^2$ we can always set $q := \chi^2_{n, \gamma}$, as this calibrates the confidence set from the observation space.
Under the linearity assumption in \Cref{eq:linear_model} when $\bK$ is full column rank, we can tighten the constraint to $q := \lVert \tilde{\by} - \bK \hat{\btheta} \rVert_2^2 + \chi^2_{p, \gamma}$, where $\hat{\btheta}$ is the maximum likelihood estimator of the model parameter vector.
This bound is the result of calibrating the confidence set to the parameter space and results in the following,
\begin{equation} \label{eq:feas_known_variance}
    D\left(\tilde{\by}, \lVert \tilde{\by} - \bK \hat{\btheta} \rVert_2^2 + \chi^2_{p, \alpha}\right) = \left\{\btheta \in \Theta : S^l(\btheta) - S^l(\hat{\btheta}) \leq \chi^2_{p, \alpha} \right\},
\end{equation}
where $S^l(\btheta) := \lVert \tilde{\by} - \bK \btheta \rVert_2^2$.
As articulated in \cite{rust_burrus_1972, rust_oleary_1994, draper1998}, \Cref{eq:feas_known_variance} can equivalently be written as an ellipsoid centered around the MLE.
The form of \Cref{eq:feas_known_variance} gives a clue about how to modify the set to accomodate the unknown variance.
Namely, \cite{draper1998} points out the following distribution,
\begin{equation} \label{eq:feas_unknown_variance}
    \frac{\left(S^l(\btheta) - S^l(\hat{\btheta} ) \right)/p}{S^l(\hat{\btheta} ) / (n - p)} \sim F(p, n - p),
\end{equation}
where $F(p, n - p)$ denotes an $F$ distribution with $n$ and $n - p$ degrees of freedom.
With some algebraic manipulation, we arrive at the following valid confidence set,
\begin{equation} \label{eq:conf_set_unknown_var}
    C_\gamma(\tilde{\by}) := \left\{\btheta : S^l(\btheta) - S^l(\hat{\btheta}) \leq S^l(\hat{\btheta} ) \frac{p}{n - p} F(p, n - p; 1 - \alpha)  \right\}.
\end{equation}
\Cref{eq:feas_unknown_variance} implies that by setting $q := S^l(\hat{\btheta}) + S^l(\hat{\btheta}) \frac{p}{n - p}F(p, n - p; 1 - \alpha)$, $D(\tilde{\by}, q)$ is a valid $1 - \gamma$ confidence set.
\Cref{eq:conf_set_unknown_var} can also be written as an ellipsoid, but with the crucial difference that its principal axes lengths are random, whereas those of \Cref{eq:feas_known_variance} are not.

% In the nonlinear setting, \Cref{eq:conf_set_unknown_var} can technically still be used to produce a valid confidence set.
% It is usually referred to as the ``lack-of-fit'' method \citep{donaldson1987, seber1989}.

\subsection{A baseline comparison: least--squares interval} \label{sec:ls_baseline}
To help articulate why natively including parameter constraints into the interval estimator as described in the previous sections is an advantageous approach, we compare against the intervals that result from taking advantage of the sampling distribution of the unconstrainted least--squares estimator.
Namely, we define an alternative approach to construct a confidence interval in \Cref{alg:two_step_proc} starting after step two.

Let $\hat{\btheta}_{LS}$ denote the parameter setting minimizing $S^l(\btheta)$ over all $\btheta \in \mathbb{R}^p$.
From classical linear--Gaussian theory, we have the following distributional result,
\begin{equation}
    \frac{\bh^T \btheta^* - \bh^T \hat{\btheta}_{LS}}{se(\bh^T \hat{\btheta}_{LS})} \sim t_{n - p}, \quad se(\bh^T \hat{\btheta}_{LS}) = s \sqrt{\bh^T (\bK^T \bK)^{-1} \bh}, \quad s = \sqrt{\frac{1}{n - p} S^l(\hat{\btheta}_{LS})},
\end{equation}
where $t_{n - p}$ denotes a Student's $t$--distribution with $n - p$ degrees of freedom \citep{seber2003}.
As a result, the confidence interval,
\begin{equation}
    I_\gamma^{LS}(\by) = \left[\bh^T \hat{\btheta}_{LS} \pm t_{n - p, \gamma} se(\bh^T \hat{\btheta}_{LS}) \right],
\end{equation}
approximately achieves the desired coverage,
\begin{equation}
    \mathbb{P} \left( \bh^T \btheta^* \in I_\gamma^{LS}(\by) \right) \approx 1 - \gamma, \quad \forall \btheta^* \in \mathbb{R}^p.
\end{equation}
Note, although the above coverage result is finite--sample when the forward model is linear, we include the approximation due to the nonlinearity.
There are asymptotic results (such as those in \cite{seber1989}) regarding the validity of such intervals.
As shown in \Cref{sec:estimating_cov_exp_len}, these baseline intervals are not valid for some tested variance levels.
One additional downside of this interval as presented is that it can include values outside of the known constraints.
As such, when comparing these interval results with those of the previous section, we intersect $I_\gamma^{LS}(\by)$ with the known bounds of the interfacial stiffness value, $\Theta_0 \subset \mathbb{R}_+$.

Although we restrict our baseline comparison to the interval construction defined above, we note that bootstrapping is an additional reasonable approach to obtaining confidence intervals for interfacial stiffness.
In our experience, the bootstrapped intervals we explored performed similiarly to the least--squares interval and as such determined their inclusion would not provide a significant contribution.
We tested three bootstrap interval constructions detailed in \cite{Davison_Hinkley_1997}; the Gaussian, the basic, and the studentized.
While the performance of these intervals was comparable to that of the least--squares interval, their construction involved substantially more computation since each simulation iteration required at least $\mathcal{O}(10^3)$ constrained nonlinear least--squares optimizations to obtain the bootstrap samples.
Similar to the least--squares interval, these methods do not guarantee an interval respecting the known parameter constraints and thus require a final step intersecting the obtained interval with the constraints.

\subsection{Propagating interfacial stiffness interval to bond strength} \label{sec:prop_interval}
Suppose a particular specimen of interest produces a vector of phase angles $\by$ as described by the data--generating model in \Cref{eq:dgm}.
Further suppose we have a collection of side data, $(x_i, z_i)$ for $i = 1, 2, \dots, N$, constituting a collection of interfacial stiffness and bond strength pairs.
Note, we use $x$ when referring to interfacial stiffness in the regression context.
We desire a bond strength interval construction, $J_\alpha(\by, \bx, \bz)$ (where $\bx$ is the vector of interfacial stiffness values and $\bz$ is the vector of bond strength values), such that for a chosen miscoverage level $\alpha \in (0, 1)$,
\begin{equation} \label{eq:prop_cov_guaran}
    \mathbb{P} \left(z^* \in J_\alpha(\by, \bx, \bz) \right) \geq 1 - \alpha,
\end{equation}
where $z^*$ is the true but unknown bond strength of the specimen.
We assume we have a procedure to compute the interfacial stiffness confidence interval and thus need an approach to propagate that interval through the regression obtained via our side data, $(x_i, z_i)_{i = 1}^N$.
We propose using a simultaneous confidence band on the inferred bond strength and interfacial stiffness relationship to find the minimum and maximum achievable values of bond strength over the given interfacial stiffness confidence interval.

One noteable advantage of this approach is its modularity.
Although we make relatively strong assumptions about both the phase angle and interfacial stiffness and bond strength pairs data--generating processes, the validity of this propagation only assumes the validity of the given interfacial stiffness interval and regression confidence band.
Thus, if one were to make different assumptions about the data--generating processes or provide more general confidence set methods, the following propagation procedure would still produce a valid confidence interval.

\subsubsection{Propagating an interfacial stiffness confidence interval through a regression} \label{subsec:prop_interval}
Suppose we have a way to obtain a valid $1 - \gamma$ confidence interval for interfacial stiffness ($I_\gamma(\by)\subset \mathbb{R}$) and that we have a valid $1 - \eta$ simultaneous confidence band on the response surface between bond strength and interfacial stiffness ($B_\eta(x) \subset \mathbb{R}$) such that,
\begin{equation} \label{eq:sim_conf_band}
    \mathbb{P} \left(\beta_1 x \in B_\eta(x), \; \; \forall x \in \mathbb{R} \right) \geq 1 - \eta.
\end{equation}
In concert with \Cref{eq:lin_relationship_bs}, \Cref{eq:sim_conf_band} assumes a linear relationship between interfacial stiffness and bond strength (i.e., $z = \beta_1 x$).
Although the statement of such a band does not require a linear assumption, it matches the proposed relationship in \cite{haldren2019}, and there are clear ways to compute $B_\eta(x)$ under a linearity assumption with Gaussian noise.

% Following Figure 9b in \cite{haldren2019}, we propose the following statistical data generating model for bond strength,
% \begin{equation} \label{eq:lin_relationship_bs}
%     z = \beta_1 \log K_0 + \varepsilon, \quad \varepsilon \sim N(0, \sigma^2).
% \end{equation}
% Assuming we have interfacial stiffness and bond strength pairs, $(x_i, z_i)_{i = 1}^N$, we regress bond strength on interfacial stiffness to obtain the maximum likelihood/least-squares estimators, $\hat{\beta}_1$ and $\hat{\beta}_0$ for a fitted model, $z = \hat{\beta}_1 x + \hat{\beta}_0$.
Note, although \Cref{eq:lin_relationship_bs} does not include an intercept term, we include the intercept term in this formulation to make the model more flexible.
Under these assumptions, we have,
\begin{equation}
    B_\eta(x) = \left[\hat{\bbeta}^T \bx \pm 2 s F_\eta(2, N - 2)\sqrt{\bx^T (\bX^T \bX)^{-1} \bx}
    % \hat{\bbeta}^T \bx + 2 s F_\eta(2, N - 2)\sqrt{\bx^T (\bX^T \bX)^{-1} \bx}  
    \right],
\end{equation}
where $\hat{\bbeta} = \begin{pmatrix}\hat{\beta}_0 & \hat{\beta}_1\end{pmatrix}^T$ are the least--squares/maximum likelihood estimators of the linear coefficients, $\bx = \begin{pmatrix}1 & x \end{pmatrix}^T$, $\bX$ is the design matrix constructed such that the $ith$ row is $\begin{pmatrix}1 & x_i \end{pmatrix}$, and $F_\eta(d_1, d_2)$ is the upper $\eta$ quantile of the $F$ distribution with $d_1$ and $d_2$ degrees of freedom.
Such an interval is a standard simultaneous response surface confidence interval, for more details see \cite{miller1981}.
According to the model in \Cref{eq:lin_relationship_bs}, the randomness in $B_\eta(x)$ is the result of measurement noise in each $z_i$.

Denote the endpoints of the interfacial stiffness interval as $I_\gamma(\by) = [x_l(\by), x_u(\by)]$ and for each $x$, denote the endpoints of the bond strength simultaneous confidence band by $B_\eta(x) = [B_\eta^l(x), B_\eta^u(x)]$.
The propagation of the interfacial stiffness interval through the simultaneous confidence band is then,
\begin{equation} \label{eq:temp_prop_int}
    J_{\gamma, \eta}(\by) := \left[B_\eta^l(x_l(\by)), B_\eta^u(x_u(\by)) \right].
\end{equation}
Note, Interval~\eqref{eq:temp_prop_int} is not yet the $J_\alpha(\by, \bx, \bz)$ interval denoted in \Cref{eq:prop_cov_guaran} as it is not yet clear how to set the miscoverage levels $\gamma, \eta$ such that the miscoverage of the final interval is $\alpha$.

% \paragraph{A key problem} We suppose we construct a $1 - \gamma$ confidence interval for interfacial stiffness and a $1 - \eta$ confidence band for the bond strength response surface as a function of interfacial stiffness as defined above.
% How should we set $\gamma, \eta \in (0, 1)$ such that the final interval is of level $1 - \alpha$?
% To begin answering this question, we produced a sketch of different scenarios that may occur in \Cref{fig:int_prop_cover_scenarios}.

\begin{figure}[h]
    \centering
    \includegraphics[width=0.85\textwidth]{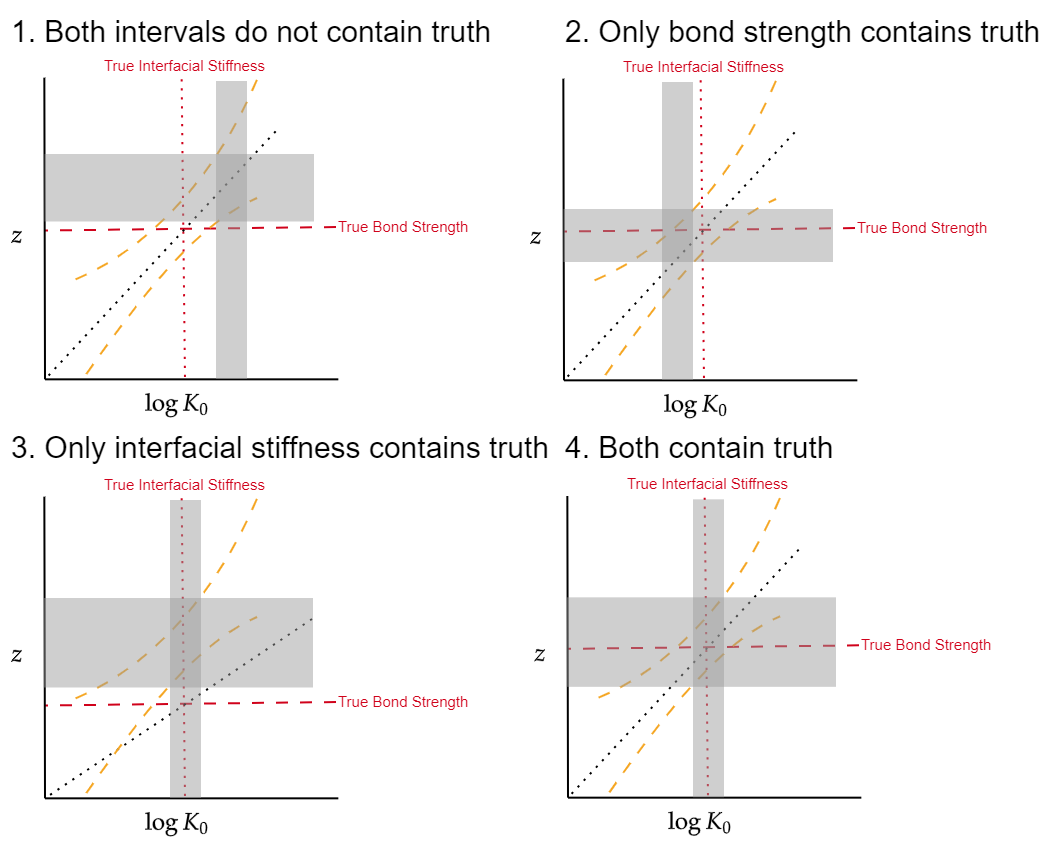}
    \caption{There are four possible scenarios that can occur when considering the coverage of both the interfacial stiffness and bond strength intervals (each interval can either cover its true value or not). The diagonal black dotted lines indicate the true relationship between interfacial stiffness ($\log K_0$) and bond strength ($z$) while the orange dashed lines indicate the simultanous confidence band ($B_\eta(x)$) resulting from the observed pairs $(x_i, z_i)_{i = 1}^N$. The gray rectangles illustrate the region defined by each respective confidence interval.}
    \label{fig:int_prop_cover_scenarios}
\end{figure}

The four possible configurations of interval coverage are shown in \Cref{fig:int_prop_cover_scenarios} to explore a few ways the propagation might be realized.
Note that the interfacial stiffness interval does not need to cover for the bond strength interval to cover, suggesting that perhaps we can set $\gamma$ such that $1 - \gamma < 1 - \alpha$.
Additionally, it appears that the only way the interfacial stiffness interval covers the true value while bond strength does not is if the simultaneous confidence band is biased.

Calibrating $J_{\gamma, \eta}(\by)$ can be challenged by the practical considerations of the experimental setup.
Namely, experimental data $(x_i, z_i)_{i = 1}^N$ are collected ahead of observing $\by$, with the simultaneous band $B_\eta(x)$ being ``amortized'' ahead of all ultrasonic observations.
If the experimental data place us in scenario three described in \Cref{fig:int_prop_cover_scenarios}, we may obtain a confidence band with poor coverage.
However, the only way this scenario can happen is if there is a systematic bias in the observations with respect to the underlying relationship.
Conversely, if the experimental data lead to any of the other scenarios, we might be able use an interfacial stiffness confidence level substantially lower than our desired level given the geometry and inherent conservatism of the simultaneous band.
We show in \Cref{app:calibration} that for a desired miscoverage level $\alpha \in (0, 1)$ for the final bond strength confidence interval, if the interfacial stiffness miscoverage level $\gamma \in (0, 1)$ and confidence band miscoverage level $\eta \in (0, 1)$ on the bond strength regression line satisfy $(1 - \gamma)(1 - \eta) = (1 - \alpha)$, then $J_{\gamma, \eta}(\by)$ is a $(1 - \alpha) \times 100$\% confidence interval.

With this relationship, once $\alpha \in (0, 1)$ has been fixed, $\gamma, \eta \in (0, 1)$ are essentially hyperparameters to tune that together parameterize an ``uncertainty budget''.
Depending on the nature of each component, it might be advantageous to spend that budget more on one confidence set than another.
For example, with $\alpha = 0.05$, we might have a substantial number of data points to build the simultaneous confidence band on the bond strength regression line and thus set $\eta = 0.005$ (since the large number of data points will drive down its size), implying that $\gamma = 0.452$.
I.e., the miscoverage level used on the interfacial stiffness interval is nearly that of the final desired miscoverage level.
If we do not have a large number of data points, we might choose $\eta = 0.045$ to shrink the confidence band size as much as possible, implying that $\gamma = 0.009$.
I.e., we ``pay'' for the shrunken confidence band size by using a smaller miscoverage level on interfacial stiffness.
In principle, it would be possible to use stochastic optimization to make these choices quantitative, but we do not pursue this idea further as we believe it would only produce marginal improvements.
There are broadly two other approaches one might pursue to construct the bond strength confidence interval.
The first is to obtain a confidence interval on interfacial stiffness and use the interval endpoints in conjunction with the raw interfacial stiffness and bond strength side data to get bounds on bond strength.
Such an approach could potentially allow for a significant relaxation in the assumed data--generating model in \Cref{eq:lin_relationship_bs}.
The second is to extend the interval optimization frameworks to work with the interfacial stiffness and bond strength data to directly optimize a confidence interval over bond strength, skipping altogether the interfacial stiffness confidence interval.
In these approaches, with infinite resources, one might consider collecting $(x_i, z_i)$ pairs \emph{after} the specimen is acquired, as the collection of data might be tailored.
Indeed, one obvious risk of the approach we have proposed is that the precollected data and amortized analysis might not contain the interfacial stiffness of the specimen.

\section{Numerical results} \label{sec:experiment}
The following section contains a collection of simulated experiments based on the data--generating models described by \Cref{eq:dgm} and \Cref{eq:lin_relationship_bs} which are used to explore the application and properties of the SSB method.
\Cref{sec:estimating_cov_exp_len} considers two $\btheta^*$ settings for \Cref{eq:dgm} (see \Cref{tab:parameters} for the ``typical'' and ``boundary'' settings) and compares SSB and least--squares interval coverage and expected length across 20 noise levels showing that constraints in the SSB interval significantly improve expected length when one of the parameters is near the constraint boundary.
\Cref{sec:bond_strength_model_effects} considers the expected length effects of the parameters defining the bond strength model in \Cref{eq:lin_relationship_bs} as these effects directly impact the usability of this method as an NDE procedure.
Finally, to demonstrate how constraints act as a proxy for information, \Cref{sec:constraint_effects} explores the effects of constraint contraction on both coverage and expected length, ultimately showing that tighter constraints lead to tighter intervals.

\begin{table}[h!]
\centering
\renewcommand{\arraystretch}{1.2} % increases row height for readability
\begin{tabular}{c c c c c c}
\toprule
Parameter ($\btheta^*$) & Units & \text{Typical} & \text{Boundary} & \text{Lower Bound} & \text{Upper Bound} \\
\midrule
$\log$ Interfacial Stiffness ($\log K_0$) & Np/m & $14.85$ & $14.85$ & $10$ & $20$ \\
Acoustic Attenuation ($\alpha_0$) & N/m$^3$ & $8.05 \times 10^3$  & $\mathbf{10^4 - 10^{-1}}$  & $0$  & $10^4$  \\
$a$ Affine Parameter & deg/Hz & $9.62 \times 10^{-6}$  & $9.62 \times 10^{-6}$  & $-3 \times 10^{-5}$   & $3 \times 10^{-5}$  \\
$b$ Affine Parameter & deg & $-42.19$  & $-42.19$ & $-100$  & $100$  \\
Bondline thickness ($L_{BL}$) & m & $9.53 \times 10^{-5}$   & $9.53 \times 10^{-5}$   & $0$    & $10^{-4}$   \\
\bottomrule
\end{tabular}
\caption{Typical and Boundary parameter ($\btheta$) settings with values and bounds used to generate the results in \Cref{sec:experiment}. Note that the ``boundary'' setting is distinguished from the ``typical'' setting only by Acoustic Attenuation ($\alpha_0$) and is set to its upper bound minus a small value. The values and bounds used here are inspired by those in \cite{haldren2019}.}
\label{tab:parameters}
\end{table}

\subsection{Estimating coverage and expected interval length} \label{sec:estimating_cov_exp_len}
We demonstrate that our proposed SSB intervals behave as expected and explore how this construction compares with the baseline least--squares interval in several scenarios.
It has previously been shown \citep{batlle2023, stanley2025} that confidence interval coverage can be most severely disrupted on or near the boundary of the parameter constraints.
As such, we use two true parameter scenarios; a ``typical'' setting where the true parameter is situated well within all constraints and a ``boundary'' setting where acoustic attenuation is set near its lower constraint while all other four parameters are shared with the ``typical'' setting (see \Cref{tab:parameters} for the details).
For each of these scenarios, we consider our interval performance over a range of noise levels ($1 \leq \sigma \leq 10$ deg) with respect to coverage and expected length by resampling $N = 2 \times 10^3$ times the data--generating process described by \Cref{eq:dgm} for the interfacial stiffness intervals and the same number of times from \Cref{eq:lin_relationship_bs} for the bond strength intervals.
Refer to the right panel of \Cref{fig:data_viz} to get an intuition for the considered noise levels.
When $\sigma = 1$, even for low frequences the signal--to--noise (SNR) ratio is large.
By contrast, when $\sigma = 10$, SNR is lower even when phase angle is at its zenith.
The uncertainty in the estimated coverage is captured by $95$\% Clopper--Pearson intervals that cover pointwise over the noise levels.
Since our method requires both a computed interfacial stiffness and bond strength interval, we show coverage and expected length for each to verify that the SSB intervals achieve at least their nominal levels.
As described in \Cref{subsec:prop_interval}, to achieve a $1 - \alpha = 0.95$ confidence interval on bond strength while using a $1 - \eta = 0.99$ confidence band for bond strength regressed on interfacial stiffness, we must compute a $1 - \gamma = (1 - \alpha)(1 - \eta)^{-1} = 0.958$ confidence interval for interfacial stiffness from the ultrasonic phase measurements.
Since simultaneous confidence bands for simple linear regression under linear--Gaussian assumptions as in \Cref{eq:lin_relationship_bs} are well--understood, we do not include any visualization of this step.

\begin{figure}[h]
    \centering
    \includegraphics[width=\textwidth]{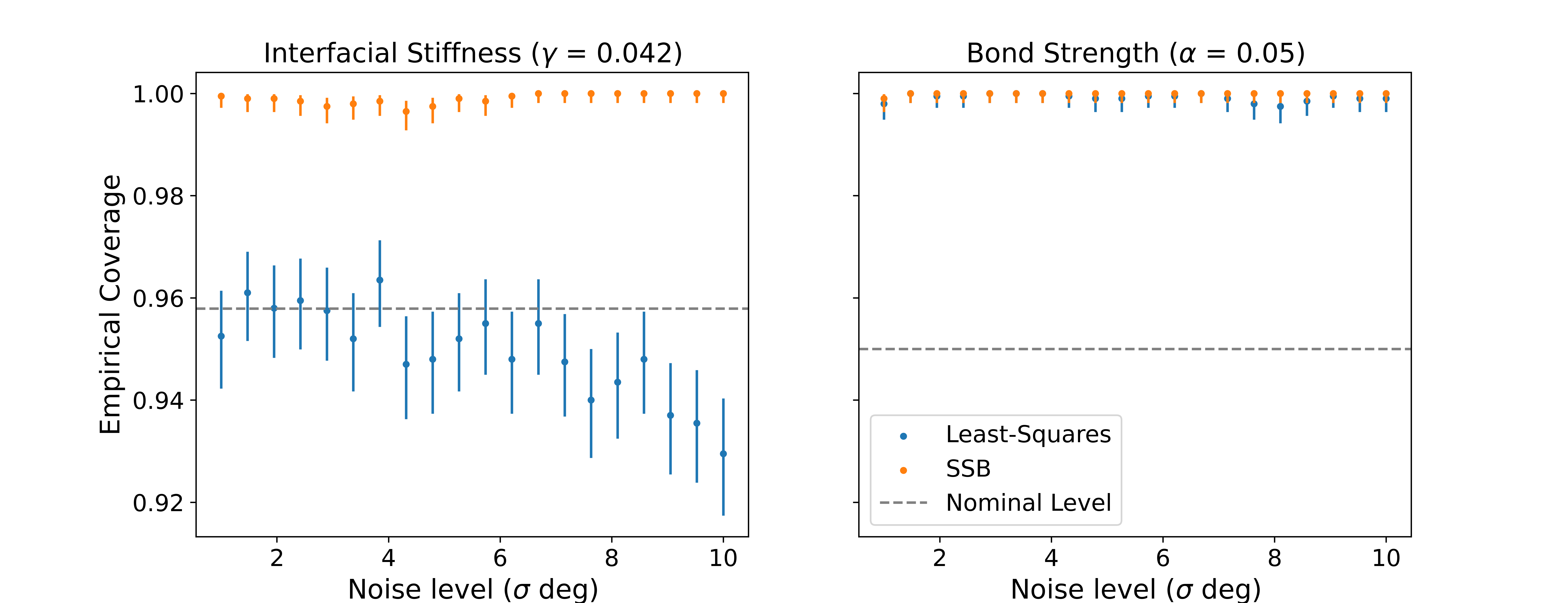}
    \caption{Coverage of both interfacial stiffness and bond strength intervals under the ``typical'' setting such that the bond strength intervals achieve at least $95$\% coverage. SSB intervals achieve their desired coverage levels ($1 - \gamma = 0.958$ for interfacial stiffness and $1 - \alpha = 0.95$ for bond strength). The least--squares intervals are included for comparison.}
    \label{fig:cov_typ_95}
\end{figure}

\begin{figure}[h]
    \centering
    \includegraphics[width=\textwidth]{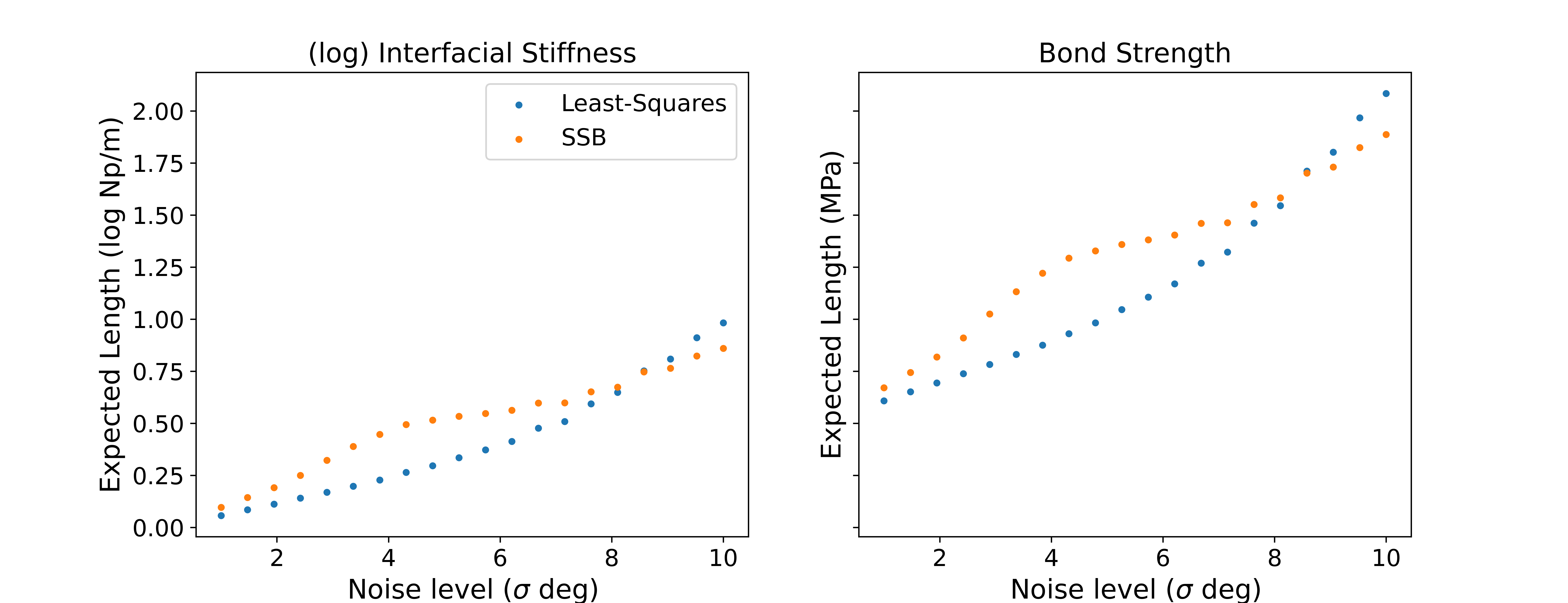}
    \caption{Expected lengths of both interfacial stiffness and bond strength intervals under the ``typical'' setting such that the bond strength intervals achieve at least $95$\% coverage. Our SSB intervals out--perform the baseline least--squares intervals at the larger tested noise levels.}
    \label{fig:exp_len_typ_95}
\end{figure}

The ``typical'' setting results are shown in Figures~\eqref{fig:cov_typ_95} and \eqref{fig:exp_len_typ_95}, showing the coverage and expected length, respectively.
The SSB intervals achieve the desired $1 - \gamma$ coverage across all noise levels while the Least--Squares intervals fail to achieve coverage in tested higher--noise regimes.
The over--coverage seen in the SSB interval coverage is expected \citep{stanley2022} since the intervals are calibrated to achieve the desired coverage level for \emph{any} possible QoI.
Both SSB and least--squares over--cover when propagating the interfacial stiffness interval to bond strength.
The extent of this over--coverage is largely driven by the slope coefficient in the relationship between interfacial stiffness and bond strength along with the variance in \Cref{eq:lin_relationship_bs}.
While both the SSB and least--squares intervals achieve the desired coverage level over all noise levels, their expected length performance varies over noise levels (see \Cref{fig:exp_len_typ_95}).
For both interfacial stiffness and bond strength, SSB achieves better length performance at higher noise levels likely because the length of the least--squares intervals is largely driven by the conditioning number of the linearized forward model, which can be large when the variance in \Cref{eq:dgm} is large.

\begin{figure}[h]
    \centering
    \includegraphics[width=\textwidth]{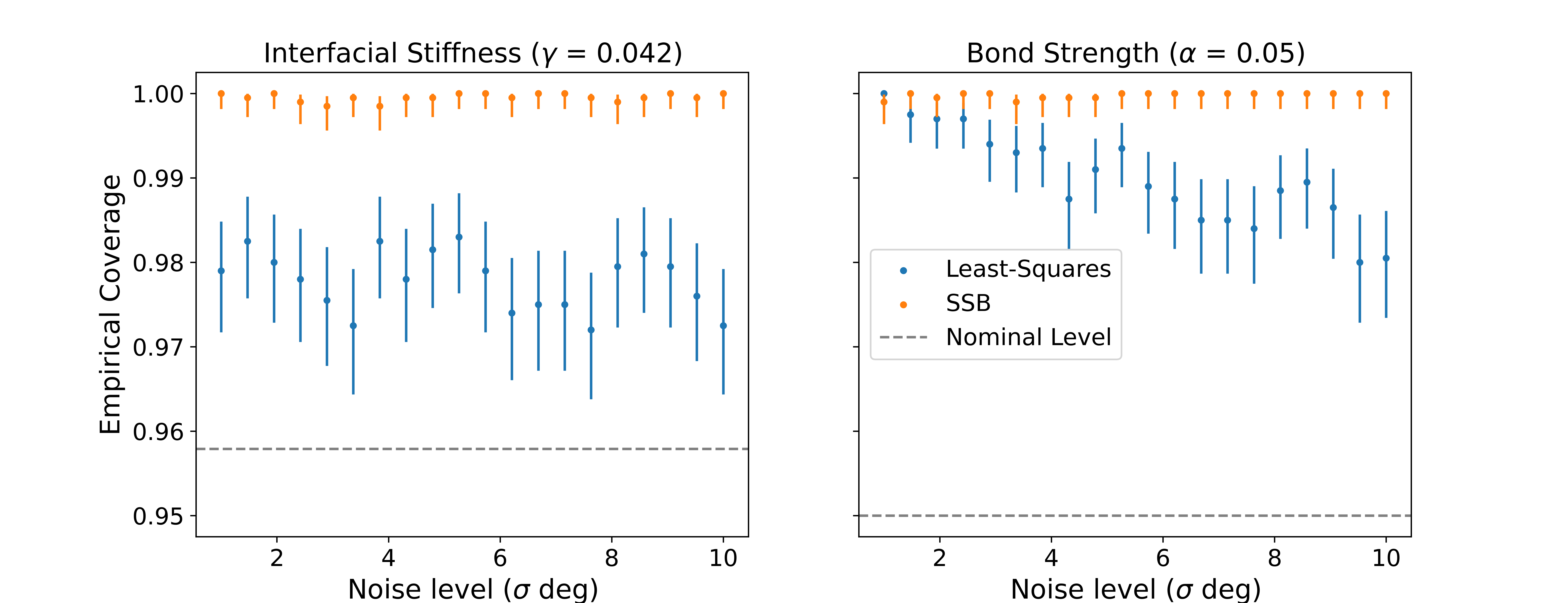}
    \caption{Coverage of both interfacial stiffness and bond strength intervals under the ``boundary'' setting such that the bond strength intervals achieve at least $95$\% coverage. SSB intervals achieve their desired coverage levels ($1 - \gamma = 0.958$ for interfacial stiffness and $1 - \alpha = 0.95$ for bond strength). The least--squares intervals are included for comparison.}
    \label{fig:cov_bd_95}
\end{figure}

\begin{figure}[h]
    \centering
    \includegraphics[width=\textwidth]{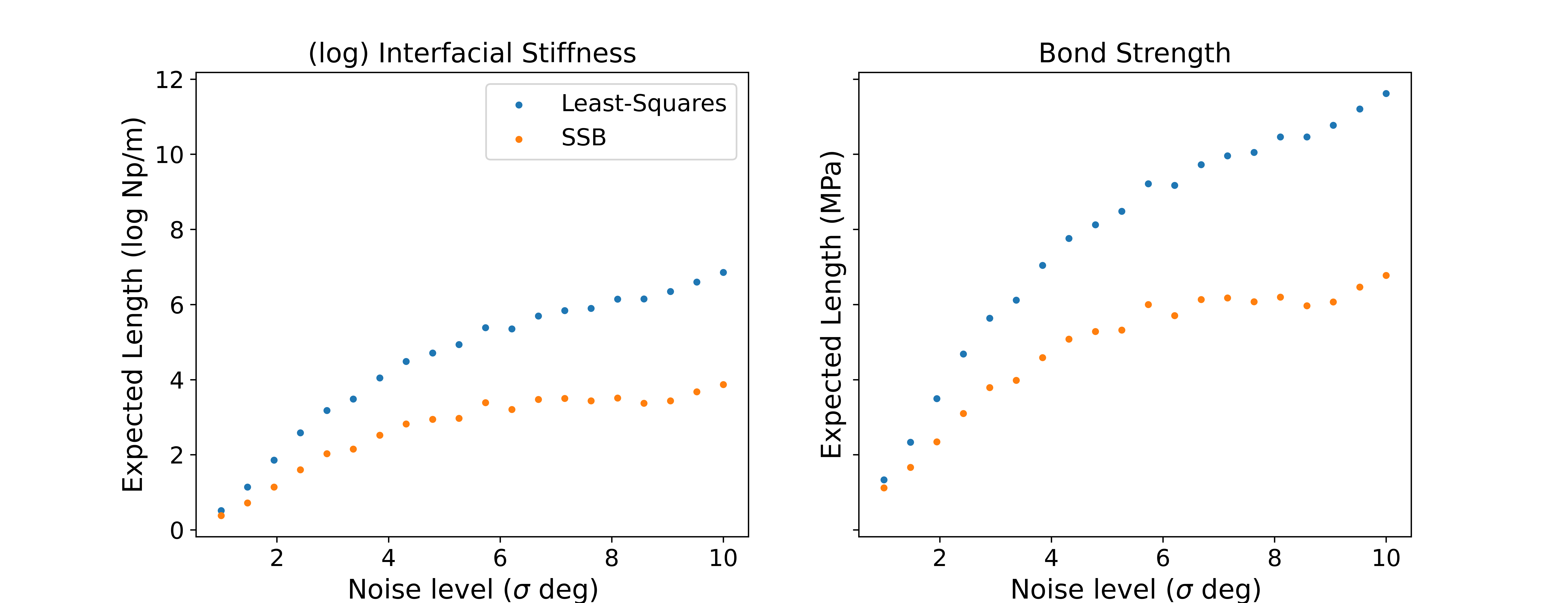}
    \caption{Expected lengths of both interfacial stiffness ($\log K_0$) and bond strength intervals under the ``boundary'' setting such that the bond strength intervals achieve at least $95$\% coverage. Our SSB intervals significantly out--perform the baseline least--squares intervals at all noise levels, showing the strong effect of nuisance parameters situated near the constraint boundary.}
    \label{fig:exp_len_bd_95}
\end{figure}

The ``boundary'' setting results are shown in Figures~\eqref{fig:cov_bd_95} and \eqref{fig:exp_len_bd_95}, showing the coverage and expected length, respectively.
Like in the ``typical'' setting, the SSB intervals over--cover for all noise levels for both interfacial stiffness and bond strength.
On the other hand, the least--squares comparison has markedly different behavior with no under--coverage for interfacial stiffness and less over--coverage for bond strength.
Although the least--squares intervals appear to out--perform the SSB with respect to coverage, they dramatically under--perform SSB with respect to expected length.
Across all noise levels, the expected lengths of the least--squares interfacial stiffness and propagated bond strength intervals are significantly larger than those of SSB.
Although setting the acoustic attenuation parameter value to be near the boundary increased the resulting expected lengths for both interval approaches, the increase is more pronounced for the least--squares approach, indirectly showing the effect of including more information in the interval construction via the parameter constraints.

\subsection{The expected length effects of bond strength model parameters} \label{sec:bond_strength_model_effects}
\begin{figure}[h]
    \centering
    \includegraphics[width=\textwidth]{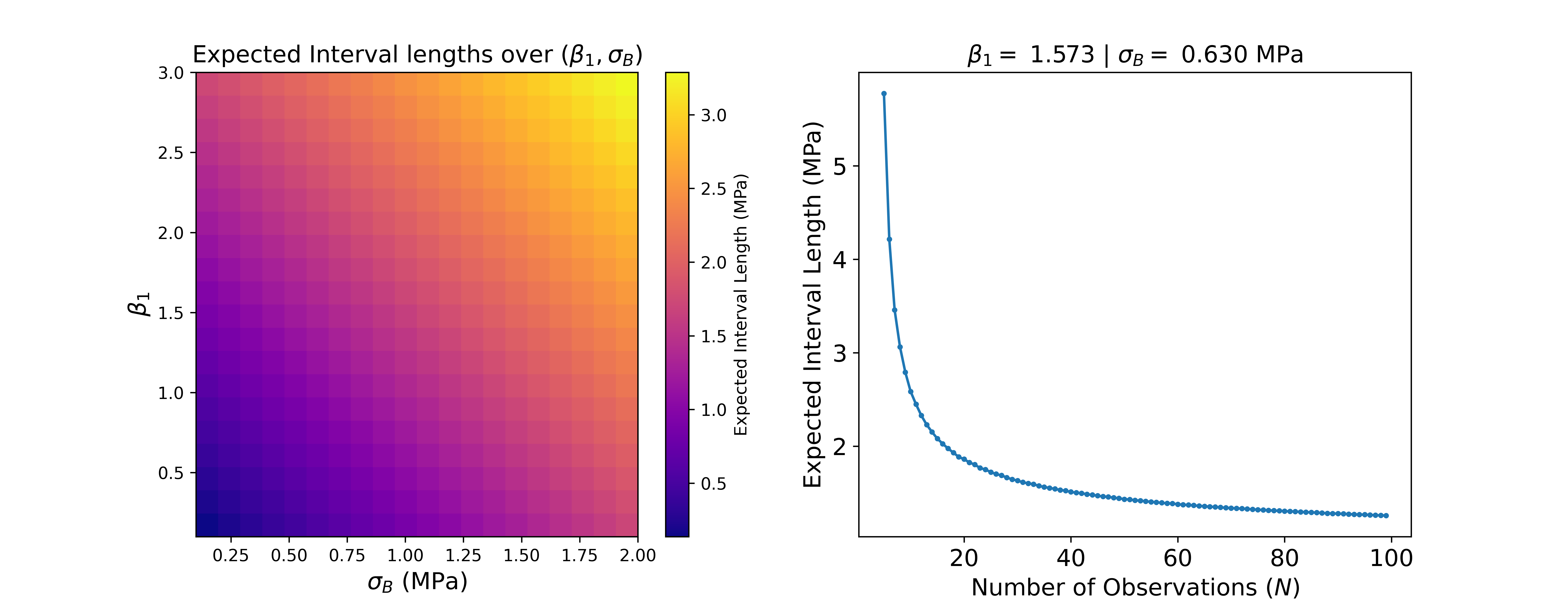}
    \caption{Expected bond strength interval lengths as a function of the parameters defining \Cref{eq:lin_relationship_bs} (i.e., $\beta_1, \sigma_B, N$). The left plot fixes $N = 60$ (the setting used elsewhere) to explore the expected interval lengths as a function of the slope and noise parameters. Similarly, the right plot fixes $\beta_1 = 1.573$ and $\sigma_B = 0.630$ to explore the expected interval length as a function of sample size.}
    \label{fig:bond_model_parameter_effects}
\end{figure}

The final propagated bond strength interval length is a function of the parameters in \Cref{eq:lin_relationship_bs} ($\beta_1, \sigma_B^2, N$) in addition to the interfacial stiffness interval.
Since we have based our simulation results in \Cref{sec:estimating_cov_exp_len} on the linear regression parameters estimated in \cite{haldren2019}, this section explores how the expected length of the propagated bond strength interval varies as a function of the linear model parameters.
The sensitivity of the expected bond strength interval length is related to the practical useability of this method, since long bond strength intervals will have a lower likelihood of providing a clear indication of safe or not safe in the context of nondestructive evaluation.
The left panel of \Cref{fig:bond_model_parameter_effects} shows a heatmap of the expected interval lengths as a function of $\beta_1$ (the coefficient of $\log$-interfacial stiffness) and $\sigma_B$ (the noise level).
The right panel of \Cref{fig:bond_model_parameter_effects} shows expected interval length as a function of the number of pairs of interfacial stiffness/bond strength observations used to fit the linear regression.
Each of the expected interval lengths shown in this section is based on the ``typical'' model parameters and the expectation is estimated using the same $2 \times 10^3$ interfacial stiffness intervals generated for \Cref{sec:estimating_cov_exp_len} along with $2 \times 10^3$ realizations of the bond strength model (\Cref{eq:lin_relationship_bs}).
Note, we alternatively could have also resampled the interfacial stiffness intervals for each bond strength model parameter settings, but since each of these settings is independent and we are only looking to estimate expected interval length, we deemed this alternative unneccesary.

For each $\log$-interfacial stiffness coefficient ($\beta_1$), there is a monotonically increasing relationship between noise ($\sigma_B$) and expected interval length.
Since the width of the simultanous confidence band of the interfacial stiffness/bond strength relationship increases with noise, this relationship is sensible.
Similarly, for each noise value, expected interval length increases monotonically with the $\log$-interfacial stiffness coefficient.
This increase is also sensible, as a larger coefficient means that the width of the confidence band has higher leverage on the bond strength interval.

While the $\log$-interfacial stiffness coefficient and noise are idealized to be fixed, unknown, and uncontrollable parameters, the number of observations ($N$) is controllable, subject to practical constraints.
Its relationship with expected interval size is highly nonlinear with expected interval length decreasing rapidly in $N$ (\Cref{fig:bond_model_parameter_effects}).
Clearly, more observations means smaller bond strength intervals, though diminishing returns begin in the $20-60$ observation range.

\subsection{Constraint effects} \label{sec:constraint_effects}
\begin{figure}[h]
    \centering
    \includegraphics[width=\textwidth]{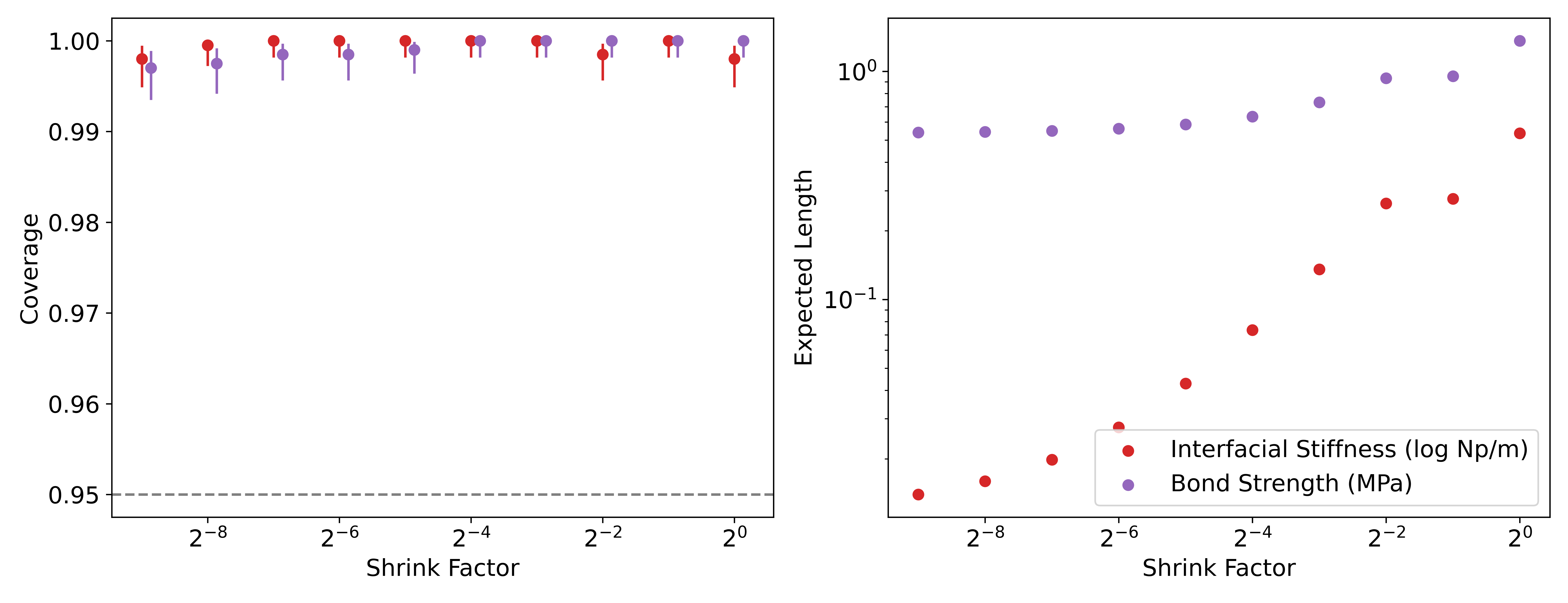}
    \caption{Coverage and expected length as a function of shrinking all parameter constraint sets aside from interfacial stiffness. When the box constraints on the parameter $\btheta$ are contracted by factors of $2$ (e.g., shrinking the box constraints by half such that the true parameter value is within the contracted interval) we are effectively adding information into the $95$\% confidence intervals and thus strong contractions (i.e., the left side of each plot) are associated with smaller intervals for both interfacial stiffness and the propagated bond strength intervals. \textbf{(Left)} Coverage levels remain approximately the same across all shrink factors. \textbf{(Right)} Expected interval length is substantially diminsihed for more substantial contractions.}
    \label{fig:constraint_effects}
\end{figure}

As seen most clearly in \Cref{fig:exp_len_bd_95}, using constraints natively in the construction of a confidence interval can substantially improve its expected length properties.
Intuitively, constraints allow us to be more precise and by extension, tight constraints allow us to be more precise than loose constraints.
We demonstrate the validity of this intuition in our simulation setting under the ``typical'' parameter setting by showing how the expected lengths of the interfacial stiffness and bond strength intervals change as we contract the original box constraints.
We consider ten sequential contractions such that the $i$th contracted interval constraint is half the length of the $(i - 1)$th contracted constraint.
We contract all parameter interval constraints aside from interfacial stiffness to simulate a setting in which the information (constraints) about the parameter of interest remains constant while the information about the other parameters increases (constraints decrease in size).
More precisely, each contracted interval constraint length is obtained by multiplying the original interval constraint length by $2^{-i}$ for $i = 0, \dots, 9$ such that for all $i$, the true ``typical'' parameter is still contained within the constraint set.
\footnote{To ensure the inclusion of the true parameter, unless an interval of the contracted length centered on the true parameter value fits entirely within the previous interval constraint, we fix the contracted interval constraint to the previous interval constraint endpoint such that the true parameter value is contained.}
To estimate coverage and expected length, we again sample \Cref{eq:dgm} and \Cref{eq:lin_relationship_bs} $2 \times 10^3$ times for all ten levels of constraint contraction.
We use the ``typical'' parameter for the forward model and the middle noise level ($\sigma = 5.74$) relative to those tried in \Cref{sec:estimating_cov_exp_len} in \Cref{eq:dgm}.
For the bond strength model, we use the same $(\beta_1, \sigma, N)$ parameters as the experiments in \Cref{sec:estimating_cov_exp_len}.
\Cref{fig:constraint_effects} shows the result of this experiment.

Across all contraction levels, we observe similar interval coverage levels both for interfacial stiffness and bond strength.
More interestingly, as the strength of the contraction increases (i.e., going from right to left in each plot in \Cref{fig:constraint_effects}) we observe an extremely rapid decrease in expected interval length primarily for interfacial stiffness.
By extension, the bond strength intervals also decrease in size, but not as rapidly in part because the decrease is attenuated by the slope characterizing the bond strength/interfacial stiffness relationship.
This exponential decrease is perhaps the best-case scenario, as it is constructed by simultaneously contracting all parameter constraints (aside from interfacial stiffness).
Although we would still see the same qualitative effect if only a subset of the parameter constraints were contracted, the effect would likely not be as pronounced.

\section{Discussion and conclusion} \label{sec:disc_conc}
% recap
Through the extensive simulated results in \Cref{sec:experiment}, we show that our proposed method is able to constrain bond strength via confidence intervals that respect the desired coverage level.
With the relaxation of the linearity and known variance assumptions of the optimization--based confidence interval, our method proposed in \Cref{sec:method} leverages known parameter constraints to produce an interval that is competitive with the baseline in a ``typical'' case and significantly superior in terms of expected length in the ``boundary'' case.
Our method's superiority in the latter case highlights the importance of controlling for nuisance parameters in statistical inference.

% future steps
The results in this paper can be extended both methodogically and experimentally.
Methodologically, we focus on ``simultaneous'' confidence intervals, where crucially the feasible set for the interval endpoint optimizations must be a valid confidence set.
As explored in \cite{batlle2023,stanley2025}, when dealing with a single functional as in this scenario, one can construct significantly tighter confidence intervals if one allows for the possibility that the feasible set is not necessarily a valid confidence set.
Applying that idea to this situation would require again handling the nonlinearity and unknown variance assumptions, which is currently unclear.
Additionally, we hypothesize that there are ways to unite the two steps of our approach such that the ultrasonic phase data and interfacial stiffness and bond strength pairs are used simultaneously to construct the bond strength interval.
Intuitively we would expect that reducing the interval construction to a single step would improve the expected length properties, but this too is unclear.
More fundamentally, this paper used the simplified model from \cite{haldren2019} instead of a full partial differential equation description of the ultrasonic wave propagation.
A finite--element model of this system could be incorporated into the interval construction to provide more sophisticated results. 
Finally, we plan to implement this method on specimens that are currently being produced in the lab.
Depending on practical constraints and data availability, it is possible that further methodological innovation is required.

\clearpage
\bibliographystyle{apalike}
\bibliography{referencesmw}
\clearpage

\appendix
\section{Mathematical Symbols} \label{app:math_symbols}
\begin{table}[ht]
\centering
\renewcommand{\arraystretch}{1.2} % adds spacing between rows
% \begin{tabular}{>{$}c<{$} l >{$}c<{$} l}
\begin{tabular}{>{$}c<{$} p{6cm} >{$}c<{$} p{6cm}}
\hline
\text{Symbol} & \text{Description} & \text{Symbol} & \text{Description} \\
\hline
\alpha \in (0, 1) & Overall miscoverage level & \by \in \mathbb{R}^n & Observed phases \\
\btheta^* \in \mathbb{R}^p & True but unknown parameter vector & \bomega \in \mathbb{R}^n & Frequencies at which phases are observed \\
f(\btheta^*; \bomega) & Forward model & \bepsilon \in \mathbb{R}^n & Noise vector \\
\Theta & Known parameter constraint set & \sigma^2 & Phase noise variance \\
\bI_n \in \mathbb{R}^{n \times n} & Identity matrix & \log K_0 & Interfacial stiffness \\
\alpha_0 & Acoustic attenuation & a, b & Affine adjustment parameters \\
L_{BL} & Bond Length (BL) & z & Bond strength \\
\beta_1 & $\log K_0$ coefficient & \sigma^2_B & Noise variance of bond strength observation \\
\hat{\beta_0}, \hat{\beta_1} & least--squares coefficients & \bK \in \mathbb{R}^{n \times p} & Linearized forward model \\
I_\alpha(\by) \subset \mathbb{R} & $1 - \alpha$ interfacial stiffness CI & D(\by, q) & Feasible set for interval optimization \\
q > 0 & Feasible set cutoff & \bh \in \mathbb{R}^p & Vector defining quantity of interest \\
\chi^2_n & Chi--squared distribution with $n$ DoF & \chi^2_{n, \alpha} & Upper--$\alpha$ quantile of $\chi^2_n$ distribution \\
\hat{\btheta} \in \Theta & Constrained nonlinear LS estimator & \gamma \in (0, 1) & Miscoverage level of interfacial stiffness interval \\
F_\gamma(p, n - p) & F--distribution upper $\gamma$ quantile with $(p, n - p)$ DoFs & \mathcal{S}(\btheta) & Squared Euclidean norm of residuals under $\btheta$ \\
\mathcal{S}^l(\btheta) & Linearized residual norm under $\btheta$ & F(p, n - p) & F--distribution with $(p, n - p)$ DoF \\
\tilde{\by} \in \mathbb{R}^n & Linearized observation $\by$ & C_\gamma(\tilde{\by}) & $1 - \gamma$ conf. set for parameter $\btheta^*$ \\
se(\bh^T \hat{\btheta}_{LS}) & Standard error of QoI estimator & t_{n - p} & $t$--distribution with $n - p$ DoF \\
(x_i, z_i) & $i$th interfacial stiffness and bond strength observation & \bx \in \mathbb{R}^N & Vector of interfacial stiffness observations \\
\bz \in \mathbb{R}^N & Vector of bond strength observations & J_\alpha(\by, \bx, \bz) \subset \mathbb{R} & $1 - \alpha$ bond strength CI \\
x^* > 0 & True interfacial stiffness & z^* > 0 & True bond strength \\
\eta \in (0, 1) & Miscoverage level of bond strength confidence band & B_\eta(x) & $1 - \eta$ bond strength confidence band at interfacial stiffness $x$ \\
\hline
\end{tabular}
\caption{List of symbols used in the paper presented in appearance order.}
\label{tab:symbols}
\end{table}

\section{Calibrating the propagated interval coverage} \label{app:calibration}
We now show how to calibrate the coverage of $J_{\gamma, \eta}(\by)$ using observed data $\by$ and $(x_i, z_i)_{i = 1}^N$.
We show below that by setting $\gamma, \eta \in (0, 1)$ such that $(1 - \gamma)(1 - \eta) = (1 - \alpha)$, we obtain a $(1 - \alpha) \times 100$\% bond strength confidence interval.
Let $x^*$ and $z^*$ denote the true interfacial stiffness and bond strength values, respectively.
By construction, if $z^* \in B_\eta(x^*)$ and $x^* \in I_\gamma(\by)$, it follows that $z^* \in J_{\gamma, \eta}(\by)$.
Using this fact and the law of total probability, we obtain the following inequalities:
\begin{align}
    \mathbb{P}\left(z^* \in J_{\gamma, \eta}(\by) \right) &= \mathbb{P} \left( z^* \in J_{\gamma, \eta}(\by), z^* \in B_\eta(x^*) \right) + \mathbb{P} \left( z^* \in J_{\gamma, \eta}(\by), z^* \notin B_\eta(x^*) \right) \nonumber \\
    &\geq \mathbb{P} \left( z^* \in J_{\gamma, \eta}(\by), z^* \in B_\eta(x^*) \right) \nonumber \\
    &= \mathbb{P} \left( z^* \in J_{\gamma, \eta}(\by), z^* \in B_\eta(x^*), x^* \in I_\alpha(\by) \right) + \mathbb{P} \left( z^* \in J_{\gamma, \eta}(\by), z^* \in B_\eta(x^*), x^* \notin I_\alpha(\by) \right) \nonumber \\
    &\geq \mathbb{P} \left( z^* \in J_{\gamma, \eta}(\by), z^* \in B_\eta(x^*), x^* \in I_\alpha(\by) \right) \nonumber  \\
    &= \mathbb{P} \left(z^* \in B_\eta(x^*), x^* \in I_\alpha(\by) \right) \nonumber \\
    &= \mathbb{P} \left(z^* \in B_\eta(x^*) \right) \mathbb{P} \left(x^* \in I_\alpha(\by) \right) \nonumber \\
    &= (1 - \eta)(1 - \gamma), \label{eq:coverage_ub}
\end{align}
where the final line follows from the statistical independence of the two events.
By setting \Cref{eq:coverage_ub} to $1 - \alpha$, we obtain the desired lower bound.
This relationship implies $(\alpha - \eta) (1 - \eta)^{-1} = \gamma$ which implies that $\alpha > \eta$, otherwise $\gamma \leq 0$.
Thus, the confidence band level cannot be lower than the desired level of the propagated interval.

\end{document}